\documentstyle[prl,preprint,aps]{revtex}

\def\AmS{{\protect\the\textfont2
        A\kern-.1667em\lower.5ex\hbox{M}\kern-.125emS}}

\makeatletter
\tighten
\begin{document}
\draft
\title{Density expansion for the mobility of electrons in Helium gas}
\author{K.I.Wysokinski}
\address{Institute of Physics,\\
 M.Curie-Sklodowska University,\\
 20-031 Lublin, Poland\\
 and\\
 Materials Science Institute\\
 University of Oregon\\
 Eugene, OR 97403}
\author{Wansoo Park and D.Belitz}
\address{Department of Physics and Materials Science Institute,\\
 University of Oregon,\\
 Eugene, OR 97403}
\author{T.R.Kirkpatrick}
\address{Institute for Physical Science and Technology,\\
 University of Maryland,\\
 College Park, MD 20742}

\date{\today}
\maketitle
\begin{abstract}
We calculate the electron mobility for a
quantum Lorentz model, which provides a realistic description of electrons
in Helium gas, to second order in the gas density. We show that this
provides sufficient theoretical information to
allow for an experimental
observation of the famous logarithmic term in the density expansion.
Detailed predictions, and a discussion of a suitable
parameter range, for such an experiment are given.
\end{abstract}

\narrowtext

\par
The absence of a virial expansion for transport coefficients
\cite{DorfmanCohen} has become famous as one of the surprises in
Theoretical Physics \cite{Peierls}. For the sake of definiteness, let us
consider the example of a tagged particle in a fluid. If one tries to
expand the diffusion coefficient, $D$, in powers of the fluid density,
$n$, for a $d$-dimensional system one encounters a nonanalyticity of
the form $n^{d-2}\ln n$. This is true for both classical
\cite{DorfmanCohen} and quantum mechanical \cite{LangerNeal} tagged
particles. An analogous term is found in other transport coefficients.
The existence of this nonanalyticity, and its relation to the long-time
tail in the velocity autocorrelation function, is well established
theoretically through many calculations for various systems and models
\cite{DorfmanSengersKirkpatrick}. It also has been seen in computer
simulations of the classical $2-d$ Lorentz model
of a particle moving in a random environment of uncorrelated
hard disk scatterers \cite{Bruin}. In real experiments,
however, the effect has so far not been convincingly observed
\cite{DorfmanSengersKirkpatrick}.

\par
Considering the fundamental importance of the theoretical prediction,
and the substantial interest it has generated over almost thirty years,
the absence of an experimental confirmation is somewhat surprising.
There are many reasons for this failure. (1) Logarithmic terms on top
of an analytic background are notoriously hard to detect. (2) The
coefficients in the density expansion of the transport coefficents of
classical fluids are known, or have been estimated,
only for hard-sphere model fluids, not
for realistic interaction potentials \cite{DorfmanSengersKirkpatrick}.
(3) These estimates indicate that in classical fluids the coefficients
of the analytic terms are in general large compared to those of the
nonanalytic ones. (4) $2-d$ systems,
in which the nonanalytic term is the leading correction to the Boltzmann
value, are hard to realize classically, while the $2-d$ quantum case is
greatly complicated by localization effects and not understood theoretically
\cite{localization}. Below we will show that $3-d$ quantum systems do not
suffer from the problems (2) - (4), and are therefore the most
promising candidates for an experimental observation of the nonanalyticity.

\par
A particularly promising system consists of electrons injected into Helium
gas of density $n$ \cite{Schwarz}. The electron-Helium scattering process
is well known, and its characteristics are convenient from a theoretical
point of view. The scattering length, $a_{s}=0.63\AA$, is positive,
and for thermal electrons the energy dependence of the scattering cross
section is negligible. Since the electrons behave quantum mechanically,
the thermal wavelength, $\lambda = (2\pi^2\hbar^{2}\beta/m)^{1/2}$, with
$\beta=1/k_{B}T$ and $m$ the electron mass, provides an additional length
scale besides $a_{s}$ and the mean Helium atom separation $n^{-1/3}$.
The leading parameter in the density expansion is $na_{s}^{2}\lambda =
\lambda/4\pi l$ \cite{KirkpatrickDorfman}, with $l=1/4\pi n a_{s}^{2}$ the
mean free path, and $a_{s}/\lambda$ serves as an additional small
parameter. The mass ratio $m_{He}/m\approx 10^{4}$
makes it a good approximation
to treat the Helium atoms as static scatterers. Finally, the low density
of the injected electrons allows one to neglect Coulomb interaction
effects between the electrons. An experiment \cite{Schwarz} which measures
the mobility of the electrons (which is related to the diffusion coefficient
by an Einstein relation) thus constitutes an almost ideal realization of a
$3-d$ quantum Lorentz model.

\par
The density expansion for the $3-d$ quantum Lorentz model has been considered
in Refs.\onlinecite{KirkpatrickDorfman,KirkpatrickBelitz}. It is
convenient to calculate the conductivity, $\sigma$,
of degenerate electrons at $T=0$,
and then to convert to the experimentally relevant finite-$T$ mobility by
means of an Einstein relation and a Kubo-Greenwood formula. The leading
terms in the expansion for $\sigma$ are,
\begin{mathletters}
\label{eqs:1}
\begin{equation}
\sigma = \sigma_{B} \biggl[ 1 + \sigma_{1} {1 \over 2k_{F}l} +
\sigma_{2log}\biggl({1 \over 2k_{F}l}\biggr)^{2} \ln\biggl({1 \over
2k_{F}l}\biggr) +
\sigma_{2}\biggl({1 \over 2k_{F}l}\biggr)^{2} +
O(a_{s}/l) + o\bigl(1/(k_{F}l)^2\bigr)\biggr]\ \ \ ,
\label{eq:1a}
\end{equation}
with
\begin{equation}
\sigma_{1} = -4\pi/3\ \ \ ,
\label{eq:1b}
\end{equation}
\begin{equation}
\sigma_{2log} = (\pi^2-4)/2\ \ \ .
\label{eq:1c}
\end{equation}
\end{mathletters}
Here $k_{F}$ is the Fermi wave number, and
$\sigma_{B}=e^{2}k_{F}^{2}l/3\pi^{2}\hbar$
is the Boltzmann conductivity with $e$ the electron charge.
$a_{s}k_{F}$ is considered small, and $o(\epsilon)$ denotes terms that
vanish faster then $\epsilon$. The coefficient $\sigma_{2}$ of the
analytic term at second order is not known.

\par
Adams {\it et al.} \cite{Adamsetal} have used Eqs.(\ref{eqs:1}) to
analyze experimental data obtained from time-of-flight measurements for
electrons in He and $\rm H_{2}$.
Their main objective was to refute the popular
misconception that $\sigma_{1}=\sigma_{2log}=0$ \cite{Mott}, which
arose from an inappropriate application of localization ideas to the
low-density regime. Ref.\onlinecite{Adamsetal} concluded that the
existing experiments give very good agreement with the value of $\sigma_{1}$
given in Eq.(\ref{eq:1b}). This success raises the question whether the same
system could be used to observe the logarithmic term. In the absence of
information about $\sigma_{2}$ this would involve measuring the conductivity
over a gas density range that is sufficient to observe the logarithmic
dependence directly. This is clearly hopeless. However, if $\sigma_{2}$
was known, then the logarithmic term would just provide a weakly density
dependent correction to it, and a sufficiently accurate experiment
{\it at fixed gas density} would be sufficient to probe the existence of
the logarithmic term.

\par
In this Letter we report a calculation of $\sigma_{2}$. We then put the
density expansion in a form that can be directly compared with experiment,
and discuss the parameter regime in which our calculation provides a
sufficiently accurate description of electrons in Helium gas. We find
that our results allow for an experimental check of the existence or
otherwise of the logarithmic term by means of a time-of-flight experiment
of the type reported in Ref.\onlinecite{Schwarz}, provided that the
experimental accuracy can be increased by about a factor of ten.

\par
The theoretical framework for our calculation is the diagrammatic
approach developed by Kirkpatrick and Dorfman
\cite{KirkpatrickDorfman}. This paper showed how to formulate the
problem in terms of standard Edwards diagrams, and identified all diagrams
that contribute to $\sigma_{1}$ and $\sigma_{2log}$ in Eqs.(\ref{eqs:1}).
We have extended this calculation by identifying and calculating all
diagrams that contribute to $\sigma_{2}$. The complete classification
and evaluation of the diagrams is rather lengthy, and will be reported
elsewhere \cite{ustbp}. Here we restrict ourselves to a few general
remarks. (1) The diagrammatic contributions to $\sigma_{2}$ can be
separated into three classes: (i) The diagrams identified in Ref.
\onlinecite{KirkpatrickDorfman} as contributing to $\sigma_{1}$ and
$\sigma_{2log}$, which were calculated in Ref.\onlinecite{KirkpatrickBelitz},
all contribute to $\sigma_{2}$ as well. (ii) Certain infinite
resummations obtained from the previous diagrams by replacing
simple impurity lines by either ordinary impurity ladders ('diffusons')
or maximally crossed ladders ('Cooperons') contribute to $\sigma_{2}$.
(iii) A number ($\sim 25$) of new skeleton diagrams with up to four
impurity lines also contribute to $\sigma_{2}$. (2) Since the electron
self energy becomes momentum dependent at $O(n^{2})$ it is most convenient
to use the Green function in self-consistent Born approximation to
construct the diagrams, as was done in Ref.\onlinecite{KirkpatrickDorfman},
and to include the higher order self energy contributions explicitly.
The real part of the self-consistent Born self energy, which was
neglected in Ref.\onlinecite{KirkpatrickDorfman}, is a constant which
strictly renormalizes the chemical potential and can be neglected for
our purposes as well. (3) If one works to lowest order in the small
parameter $a_{s}k_{F}$ one encounters logarithmic singularities
signalizing a $\ln(a_{s}k_{F})$ dependence. It is therefore necessary
to use an ultraviolet cutoff momentum $Q\sim 1/a_{s}$ in certain
integrals. The resulting $\ln Q$ dependences are all due to the real
part of the self energy. They constitute shifts of the chemical
potential, and disappear if one considers the experimentally relevant
mobility instead of the conductivity.

\par
{}From our calculation we obtain,
\begin{mathletters}
\label{eqs:2}
\begin{equation}
\sigma_{2} = 4 \ln(Q/k_{F}) + {55 \over 36}\pi^{2} - 14\ln 2 +
7 - I_{1} + I_{2} - I_{3}\ \ \ ,
\label{eq:2a}
\end{equation}
where $I_{1,2,3}$ are three integrals which we could not reduce to
tabulated ones,
\begin{equation}
I_{1} = \int_{0}^{1} {dx \over x} \biggl[\ln\biggl({1-x \over 1+x}\biggr)
\biggr]^{2} = 4.207\dots\ \ \ ,\nonumber
\end{equation}
\begin{equation}
I_{2} = \int_{0}^{1} dx\ x\biggl[\ln\biggl({1-x \over 1+x}\biggr)
\biggr]^{2} = 2.772\dots\ \ \ ,
\label{eq:2b}
\end{equation}
\begin{equation}
I_{3} = {4 \over \pi} \int_{0}^{\infty} {dx \over x^{2}}
(\arctan x)^{4}\Big/\biggl[1 - {1 \over x} \arctan x\biggr]
= 7.716\dots\ \ \ .\nonumber
\end{equation}
\end{mathletters}
For the density of states a much simpler calculation yields,
\begin{mathletters}
\label{eqs:3}
\begin{equation}
N = N_{0}\biggl[1 + N_{2}\biggl({1 \over 2k_{F}l}\biggr)^{2} +
O\Bigl(1/(k_{F}l)^{3}\Bigr)\biggr]\ \ \ ,
\label{eq:3a}
\end{equation}
with $N_{0}=k_{F}m/\pi^{2}\hbar^{2}$ the free electron density of states,
and
\begin{equation}
N_{2} = 2 \ln (Q/k_{F}) - 4 \ln 2 - 1/2\ \ \ .
\label{eq:3b}
\end{equation}
\end{mathletters}

\par
We next convert these results into the temperature dependent mobility,
which is directly measured in a time-of-flight experiment. The
mobility is given by $\mu(T)=\sigma(T)/en(T)$, where $\sigma(T)$
is the temperature dependent conductivity, and $n(T)$ is the electron
particle number density. $\sigma(T)$ is obtained from Eq.(\ref{eq:1a})
by means of the Kubo-Greenwood formula \cite{Greenwood},
\begin{equation}
\mu(T) = \int_{0}^{\infty} d\epsilon \biggl({-\partial f \over
\partial \epsilon}\biggr) \sigma(\epsilon)\bigg/e\int_{0}^{\infty}
d\epsilon f(\epsilon) N(\epsilon)\ \ \ ,
\label{eq:4}
\end{equation}
with $\sigma(\epsilon)$, $N(\epsilon)$ from Eqs.(\ref{eqs:1}),
(\ref{eqs:3}) as functions of $\epsilon = \hbar^{2}k_{F}^{2}/2m$,
and $f(\epsilon)$ the Fermi function. We are interested in the limit
of small electron density where the latter can be replaced by a
Boltzmann distribution, $f(\epsilon)= e^{\beta \mu}e^{-\beta \epsilon}$.
Performing the integrals we obtain,
\begin{mathletters}
\label{eqs:5}
\begin{equation}
\mu(T)/\mu_{B}(T) = 1 + \mu_{1}\chi + \mu_{2log}\chi^{2}\ln\chi +
\mu_{2}\chi^{2} + O(\chi a_{s}/\lambda) + o(\chi^{2})\ \ \ ,
\label{eq:5a}
\end{equation}
with $\chi = \lambda/\pi l =
8\pi na_{s}^{2}(\hbar^{2} \beta/2m)^{1/2}$, and
$\mu_{B}(T) = 4el/3(2\pi m/\beta)^{1/2}$ the Boltzmann value for the
mobility. The coefficients are,
\begin{equation}
\mu_{1} = -\pi^{3/2}/6\ \ \ ,
\label{eq:5b}
\end{equation}
\begin{equation}
\mu_{2log} = (\pi^{2} - 4)/32\ \ \ ,
\label{eq:5c}
\end{equation}
\begin{equation}
\mu_{2} = {1 \over 16}\biggl[
{\pi^{2} \over 36}(55+9C) - C + 8 - 10\ln 2
-I_{1} + I_{2} - I_{3}\biggr] - 2\mu_{2log}\ln 2 = 0.236\dots\ \ \ ,
\label{eq:5d}
\end{equation}
\end{mathletters}
where $C$ is Euler's constant.

\par
Let us discuss the relation of our result, Eqs.(\ref{eqs:5}), to
existing \cite{Schwarz} and possible future experiments. The main
uncertainties in this relation arise from the terms of
$o(\chi^{2})$ and $O(\chi a_{s}/\lambda)$ in Eq.(\ref{eq:5a}).
The former are undoubtedly nonanalytic, but their functional form is
not known, let alone their magnitude. However, to estimate their
importance it is plausible to neglect the nonanalyticity, and to
assume that the coefficients in the $\chi$-expansion are all of
roughly the same magnitude. In the Kubo-Greenwood integration,
Eq.(\ref{eq:4}), the $\chi^{3}$-term picks up an extra factor of
$\sqrt{\pi}$, and so we estimate $o(\chi^{2})\approx \mu_{3}\chi^{3}$
with $-2\sqrt{\pi}\mu_{2}\alt\mu_{3}\alt 2\sqrt{\pi}\mu_{2}$. We neglect
the, presumably weak, $\chi$-dependence of $\mu_{3}$, and allow for a
safety margin in the form of an extra factor of $2$.
The terms of $O(\chi a_{s}/\lambda)$ describe
excluded volume-like effects, {\it i.e.}
they correspond to terms of $O(na_{s}^{3})$ in the zero temperature
perturbation theory. Since the temperature dependence of these terms
is different from that of the leading terms it would in principle
be possible to separate them experimentally. However, as long as
they are small compared to the second order terms which we keep in
Eq.(\ref{eq:5a}), {\it i.e.} as long as $\chi^{2}>\chi a_{s}/\lambda$,
or $\chi>a_{s}/\lambda$, they can be neglected.
This means there is a window of $\chi$-values
for which the excluded volume terms are negligible, but the higher
order in $\chi$-terms are not yet important.

\par
We now define, as a convenient quantity directly comparable with
experiment,
\begin{mathletters}
\label{eqs:6}
\begin{equation}
f(\chi) \equiv \Bigl[\mu(T)/\mu_{B} - 1
- \mu_{1}\chi\Bigr]/\chi^{2}\ \ \ .
\label{eq:6a}
\end{equation}
Our theoretical prediction for this quantity is,
\begin{equation}
f(\chi) = \mu_{2log} \ln \chi + \mu_{2} \pm \mu_{2} 2\sqrt{\pi} \chi
\ \ \  ,
\label{eq:6b}
\end{equation}
\end{mathletters}
with $\mu_{2}$ and $\mu_{2log}$ from Eqs.(\ref{eqs:5}). We have
omitted the error due to the excluded volume effects, since they
turn out to be negligible in the parameter range that is of experimental
interest. We now consider the experimental results obtained by
Schwarz \cite{Schwarz} at Helium temperature. At $T=4.2{\rm K}$, a
He gas density $n=10^{21}cm^{-3}$ corresponds to $\chi=1$, and data
were obtained for $\chi$ as low as $0.08$. In Fig.\ref{fig:1} we show
the theoretical prediction, Eqs.(\ref{eqs:6}), for $0<\chi<0.7$
together with Schwarz's data. The error bars shown assume a total
error of $3\%$ in $\mu/\mu_{B}$ and $4\%$ in $\chi$. To illustrate the
effect of the logarithmic
term the figure also shows what the theoretical
prediction would be if $\sigma_{2log}$ in Eqs.(\ref{eqs:1}) was zero.

\par
{}From Fig.\ref{fig:1} we draw the following conclusions. (1) The
existing data are certainly consistent with the existence of the
logarithmic term, but are not accurate enough to be conclusive.
(2) A repetition of this experiment in the region $0.1<\chi<0.2$
with an accuracy improved by at least a factor of $10$ would be
sufficient for a convincing test of the logarithmic term's existence.
This range of $\chi$-values is particularly suitable because excluded
volume effects are negligible \cite{vanLeeuwenWeyland}.
At larger $\chi$-values the uncertainty
due to the $\chi^{3}$-terms makes the theoretical prediction
meaningless, and at lower $\chi$-values errors in determining $\chi$
translate into very large errors in $f(\chi)$. Also, the excluded
volume effects become noticable at smaller $\chi$.

\par
Let us finally discuss other error sources that are due to
idealizations in the theoretical model. (1) Static correlations
between the He atoms. This effect can be estimated, {\it e.g.}, from
Ziman's formula \cite{Ziman}. It is of the same order as the excluded
volume effects discussed above. (2) Coulomb interaction effects.
Ref.\cite{Schwarz} used an electron current density
$j\approx 10^{-12}Ccm^{-2}s^{-1}$. With a drift velocity
$v\approx 10^{4}cm/s$ this
corresponds to an electron density $n_{e} \approx 10^{3}cm^{-3}$.
The Coulomb energy $E_{c}\approx e^{2}/n_{e}^{-1/3}$
is then less than one percent
of the kinetic energy, $k_{B}T$.
Obviously, Coulomb effects can be made even smaller by
decreasing the electron density.
(3) Dynamics of the He gas. During a scattering time
$\tau \approx 10^{-12}s$ the thermal velocity of the He atoms leads
to a displacement $d\approx 1\AA$, which is comparable with the
scattering length. The effect on the electron mobility is of order
$d/\lambda \sim a_{s}/\lambda$, which is of the same order as the
excluded volume effects.

\par
In summary, we have calculated the coefficient of the analytic term
at second order in the density expansion for the mobility in a
quantum Lorentz model. We have argued that this model yields a
realistic description of electrons in He gas, and that our
calculation provides the information necessary for a convincing
experimental observation of the logarithmic term in the density
expansion. A discussion of the experimental
accuracy necessary to achieve this, and of a suitable parameter
range, has been given.

\par
We gratefully acknowledge helpful discussions with P.W.Adams and
J.V.Sengers. One of us (KIW) is grateful for the warm hospitality
extended to him at the University of Oregon.
This work was supported by the NSF under
grant numbers DMR-92-17496 and DMR-92-09879.

\begin{figure}
\caption{The reduced mobility $f$, as defined in
 Eq.(\protect\ref{eq:6a}),
 vs. the density parameter $\chi=\lambda/\pi l$. The theoretical
 prediction is for $f$ to lie between the two solid lines. The
 experimental data are from Fig.9 of Ref.
 \protect\cite{Schwarz} with
 error bars estimated as described in the text. The broken lines
 show what the theoretical prediction would be in the absence of
 the logarithmic term in the density expansion.}
\label{fig:1}
\end{figure}

\end{document}